\documentclass[aps,prl,twocolumn,superscriptaddress,showpacs]{revtex4}

\usepackage{graphicx}

\begin{document}

\title{Maximum Confidence Quantum Measurements}

\author{Sarah Croke}
\email{sarah@phys.strath.ac.uk}
\affiliation{Department of Physics, University of Strathclyde, Glasgow G4 0NG, UK}
\affiliation{Department of Mathematics, University of Glasgow, Glasgow G12 8QW, UK}
\author{Erika Andersson}
\author{Stephen M. Barnett}
\affiliation{Department of Physics, University of Strathclyde, Glasgow G4 0NG, UK}
\author{Claire R. Gilson}
\affiliation{Department of Mathematics, University of Glasgow, Glasgow G12 8QW, UK}
\author{John Jeffers}
\affiliation{Department of Physics, University of Strathclyde, Glasgow G4 0NG, UK}

\pacs{03.65.Ta, 03.67.Hk}

\date{\today}

\newcommand{\bra}[1]{\langle #1|}
\newcommand{\ket}[1]{|#1\rangle}
\newcommand{\braket}[2]{\langle #1|#2\rangle}

\begin{abstract}
We consider the problem of discriminating between states of a specified set with maximum confidence.  For a set of linearly independent states unambiguous discrimination is possible if we allow for the possibility of an inconclusive result.  For linearly dependent sets an analogous measurement is one which allows us to be as confident as possible that when a given state is identified on the basis of the measurement result, it is indeed the correct state.
\end{abstract}

\maketitle

One of the problems in exploiting the capability of a quantum system for carrying information is the difficulty in extracting the information encoded in a quantum state.  It is not possible simply to measure the state of a quantum system in a single shot measurement, as the state is not itself an observable.  Thus, without some prior knowledge, the state cannot be determined with certainty and without error.  In fact this is the case unless the state is known to be one of a mutually orthogonal set.
In quantum communications, however, the receiving party has to discriminate between a known set of states $\{\hat \rho_{i}\}$ with known prior probabilities $p_{i}$ \cite{che1}.  In general, the states will not be orthogonal so that perfect discrimination is not possible and we have to settle for the best that can be done.  This means optimising a figure of merit, with the simplest being to minimise the probability of incorrectly identifying the state.  Necessary and sufficient conditions which the operators describing this minimum error measurement must satisfy are known \cite{hol,yue}, but the optimal measurement itself is known only in certain special cases \cite{yue,hel,ban,bar,eld,and}.  A second possibility, unambiguous discrimination, is possible between two non-orthogonal states if we are prepared to accept the possibility of an inconclusive result \cite{iva,die,per1}.  When the inconclusive result is not obtained, it is possible to identify the initial state with certainty.  This strategy is optimised by minimising the probability of obtaining an inconclusive result \cite{jae}.  Unambiguous discrimination can be extended to higher dimensions \cite{per2}, but it is only applicable to sets of linearly independent states \cite{che2}.  Other figures of merit include the mutual information shared by the transmitting and receiving parties \cite{davies,sasaki} and the fidelity between the state received and one transmitted on the basis of the measurement result \cite{fidelity,hunt}.  Examples of optimal minimum error, mutual information and unambiguous discrimination measurement strategies have been demonstrated in experiments on optical polarisation \cite{riis,clarke1,hut,cla,mizuno,barQI}.

For linearly dependent states the analogue of unambiguous discrimination would be a measurement which allows us to be as confident as possible that the state we infer from our measurement result is the correct one.  We take this criterion as the basis of maximum confidence measurements.  A related problem was considered by Kosut \emph{et al.} \cite{kos}, who posed the question `if the detector declares that a specific state is present, what is the probability of that state actually being present?'.  That work used a worst-case optimality criterion, i.e. considered the measurement which maximises the smallest possible value of this probability for a given set of states.  Here we consider the construction of a measurement which achieves the maximum possible value of this probability for each state in a set.  In order to do this we sometimes have to accept the possibility of an inconclusive outcome, just as we need to for unambiguous discrimination.

Any measurement can be described mathematically by a probability operator measure (POM) \cite{hel}, also known as a positive operator valued measure \cite{per3}.  Each possible measurement outcome $\omega_{i}$ is associated with a probability operator, or POM element $\hat{\Pi}_{i}$.  In order to form a physically realisable measurement, these elements must satisfy the conditions
\begin{equation}
\begin{array}{ccl}
\hat{\Pi}_{i} &\geq 0& \\
\sum_{i} \hat{\Pi}_{i} &=& \hat {\rm I}.
\end{array}
\label{POM}
\end{equation}
The probability of obtaining outcome $\omega_{j}$ as a result of measurement on a system in state $\hat{\rho}$ is given by ${\rm Tr}(\hat{\rho} \hat{\Pi}_{j})$.

Suppose that a measurement is made on a quantum system known to have been prepared in one of $N$ possible states $\{\hat{\rho}_{i}\}$, with associated a priori probabilities $\{p_{i}\}$.  Suppose further that the outcome of the measurement, denoted $\omega_{j}$, is taken to imply that the state of the system was $\hat{\rho}_{j}$.  No restrictions are placed on the number or interpretation of other possible outcomes, for the moment we are concerned only with outcome $\omega_{j}$.  How confident can we be that this outcome leads us to correctly identify the state prepared?  The quantity of interest is the probability that the prepared state was $\hat{\rho}_{j}$, \emph{given} that the outcome $\omega_{j}$ was obtained, that is $P(\hat{\rho}_{j}|\omega_{j})$.  Using Bayes Rule we can write
\begin{equation}
P(\hat{\rho}_{j}|\omega_{j}) = \frac{P(\hat{\rho}_{j}) P(\omega_{j}|\hat{\rho}_{j})}{P(\omega_{j})}
 = \frac{p_{j} {\rm Tr}(\hat{\rho}_{j} \hat{\Pi}_{j})}{{\rm Tr}(\hat{\rho} \hat{\Pi}_{j})}
\label{confidence}
\end{equation}
where $\hat{\rho} = \sum_{i} p_{i} \hat{\rho}_{i}$ is the \emph{a priori} density operator for the system.  By maximising $P(\hat{\rho}_{j}|\omega_{j})$ with respect to the probability operator $\hat{\Pi}_{j}$, we can put a limit on how well the state $\hat{\rho}_{j}$ can be identified from the others in the set.

The process of maximising $P(\hat{\rho}_{j}|\omega_{j})$ is greatly facilitated by means of the ansatz
\begin{equation}
\hat{\Pi}_{j} = c_{j} \hat{\rho}^{-1/2} \hat{Q}_{j} \hat{\rho}^{-1/2},
\label{ansatz}
\end{equation}
where $\hat{Q}_{j}$ is a positive, trace 1 operator, and thus the weighting factor $c_{j} \geq 0$ represents the probability of occurrence of outcome $\omega_{j}$, $P(\omega_{j})$.  Hence
\begin{equation}
\begin{array}{ccl}
P(\hat{\rho}_{j}|\omega_{j}) &=& p_{j} {\rm Tr}(\hat{\rho}^{-1/2} \hat{\rho}_{j} \hat{\rho}^{-1/2} \hat{Q}_{j}) \\
&=& p_{j} {\rm Tr}(\hat{\rho}_{j} \hat{\rho}^{-1}) {\rm Tr}(\hat{\rho}_{j}^{\prime} \hat{Q}_{j}),
\label{transform}
\end{array}
\end{equation}
where $\hat{\rho}_{j}^{\prime} = \hat{\rho}^{-1/2} \hat{\rho}_{j} \hat{\rho}^{-1/2}/{\rm Tr}(\hat{\rho}_{j} \hat{\rho}^{-1})$.  The operators $\hat{\rho}_{j}^{\prime}$ and $\hat{Q}_{j}$ are both positive, with unit trace, and can be thought of as density operators.  It follows, therefore, that $P(\hat{\rho}_{j}|\omega_{j})$ is maximised if $\hat{Q}_{j}$ is a projector onto the pure state that has largest overlap with $\hat{\rho}_{j}^{\prime}$:
\begin{equation}
\hat{Q}_{j} = \ket{\lambda_{j}^{\prime max}} \bra{\lambda_{j}^{\prime max}}
\label{Pj}
\end{equation}
where $\ket{\lambda_{j}^{\prime max}}$ is the eigenket of $\hat{\rho}_{j}^{\prime}$ corresponding to the largest eigenvalue $\lambda_{j}^{\prime max}$. The limit is then given by
\begin{equation}
[P(\hat{\rho}_{j}|\omega_{j})]_{max} = p_{j} {\rm Tr}(\hat{\rho}_{j} \hat{\rho}^{-1}) \lambda_{j}^{\prime max}
\label{pcorr}
\end{equation}
and is realised by the POM element
\begin{equation}
\hat{\Pi}_{j} = c_{j} \hat{\rho}^{-1/2} \ket{\lambda_{j}^{\prime max}} \bra{\lambda_{j}^{\prime max}} \hat{\rho}^{-1/2}.
\label{MCPOM}
\end{equation}
If the state $\hat \rho_j$ is pure then this simplifies to
\begin{equation}
\hat{\Pi}_{j} \propto \hat{\rho}^{-1} \hat{\rho}_{j} \hat{\rho}^{-1}.
\label{MCPOMpure}
\end{equation}
As multiplying the POM element by a constant has no effect on the expression in Eq. (\ref{confidence}), we have some freedom in choosing the constants of proportionality and each choice will correspond to a distinct maximum confidence strategy.  In some cases it will be possible to choose the $c_j$ such that we can form a complete measurement from operators independently optimised in this way.  In other cases an inconclusive outcome is necessary, and the constants may be chosen, for example, to minimise the probability of occurrence of the inconclusive outcome.

\begin{figure}
\includegraphics[width=80truemm]{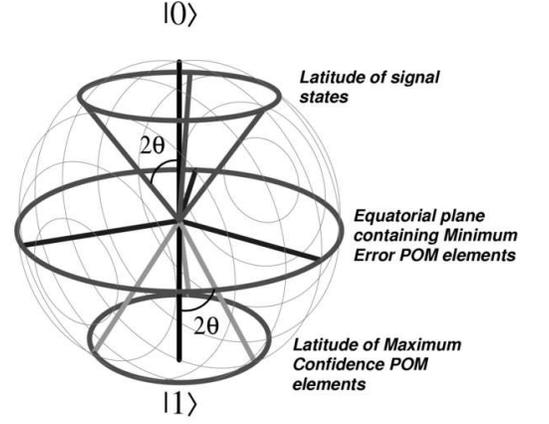}
\caption{\label{bloch} Bloch sphere representation of states.  Any density operator of a two-level system can be written $\hat{\rho} = \frac{1}{2}(\hat {\rm I} + \mathbf{r} \cdot \mathbf{\hat{\sigma}})$ where $\mathbf{r}$ is a real 3 component vector, $\mathbf{\hat{\sigma}} = (\hat{\sigma}_{x},\hat{\sigma}_{y},\hat{\sigma}_{z})$ and $\hat{\sigma}_{x (yz)}$ are the Pauli spin operators.  By plotting the vector $\mathbf{r}$, states of a 2-D complex system can be represented graphically in a 3-D real system.  The signal states used in the example, as well as the directions along which the POM elements lie are illustrated here.}
\end{figure}

As an example we consider the case of three equiprobable ($p_{i} = \frac{1}{3}, ~ i = 0,1,2$) symmetric qubit states that lie on the same latitude of the Bloch sphere (see Fig. \ref{bloch}).  For pure states $\hat{\rho} = \ket{\Psi} \bra{\Psi}$ and we can describe our three states by the kets
\begin{equation}
\begin{array}{ccl}
\ket{\Psi_{0}} &=& \cos \theta \ket{0} + \sin \theta \ket{1} \\
\ket{\Psi_{1}} &=& \cos \theta \ket{0} + e^{2 \pi i /3} \sin \theta \ket{1} \\
\ket{\Psi_{2}} &=& \cos \theta \ket{0} + e^{-2 \pi i /3} \sin \theta \ket{1}
\end{array}
\end{equation}
where $\ket{0}$, $\ket{1}$ form an orthogonal basis for the qubit and, without loss of generality, we set $0\leq \theta \leq \pi/4$.  For this set of states $\hat{\rho} = \cos^{2} \theta \ket{0} \bra{0} + \sin^{2} \theta \ket{1} \bra{1}$, and the maximum confidence POM elements are easily calculated using Eq. (\ref{MCPOMpure}) to be $\hat{\Pi}_{i} = a_{i} \ket{\phi_{i}} \bra{\phi_{i}}$ ($i = 0,1,2$), where the $a_{i}$ are positive constants and
\begin{equation}
\begin{array}{ccl}
\ket{\phi_{0}} &=& \sin \theta \ket{0} + \cos \theta \ket{1} \\
\ket{\phi_{1}} &=& \sin \theta \ket{0} + e^{2 \pi i /3} \cos \theta \ket{1} \\
\ket{\phi_{2}} &=& \sin \theta \ket{0} + e^{-2 \pi i /3} \cos \theta \ket{1}.
\end{array}
\end{equation}
Our maximum confidence if outcome $\omega_{j}$ is obtained, that is the maximum probability that the state identified is correct, is the same for each possible outcome, and is calculated from Eq. (\ref{pcorr}) to be
\begin{equation}
P(\hat{\rho}_{j}|\omega_{j}) = \frac{2}{3}.
\end{equation}
The above elements do not form a POM for any choice of the constants $a_{i}$ and hence an inconclusive result is needed.  The POM element corresponding to the inconclusive outcome is given by $\hat{\Pi}_{?} = \hat {\rm I} - \hat{\Pi}_{0} - \hat{\Pi}_{1} - \hat{\Pi}_{2}$, with a probability of occurrence $P(?) = {\rm Tr}(\hat{\rho} \hat{\Pi}_{?}) = 1 - 2 (a_{0} + a_{1} + a_{2}) \cos^{2} \theta \sin^{2} \theta$.  Different choices will give competing maximum confidence strategies and we need an additional criterion to select the best of these.  One way to do this is to follow the example of unambiguous state discrimination and to minimise P(?) subject to the constraint $\hat{\Pi}_{?} \geq 0$.  As P(?) is a monotonically decreasing function of $a_{0}$, $a_{1}$, $a_{2}$, the optimal values of these parameters lie on the boundary of the allowed domain, defined by $\hat{\Pi}_{?} \geq 0$.  This leads us to choose $a_{0}$, $a_{1}$, $a_{2}$ to be
\begin{equation}
a_{0} = a_{1} = a_{2} = (3 \cos^{2} \theta)^{-1}.
\end{equation}
The POM element corresponding to the inconclusive outcome is then of the form
\begin{equation}
\hat{\Pi}_{?} = (1 - \tan^{2} \theta) \ket{0} \bra{0},
\end{equation}
which gives the inconclusive probability
\begin{equation}
P(?) = \cos 2\theta.
\end{equation}

\begin{figure}
\includegraphics[width=80truemm]{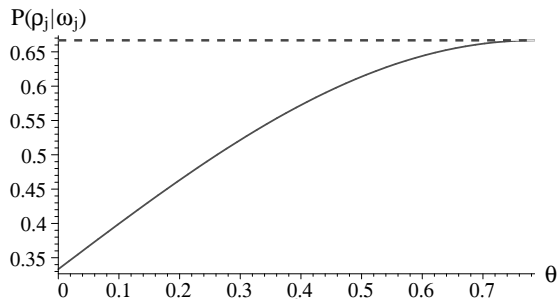}
\caption{\label{Pc} Comparison of the confidence in the state identified as a result of measurement for minimum error (solid line) and maximum confidence (dashed line) strategies.  $P(\hat{\rho}_{j}|\omega_{j})$, the probability that the state identified on the basis of measurement outcome $\omega_{j}$ is correct is plotted as a function of the parameter $\theta$.}
\end{figure}

For the purposes of comparison, we note that the minimum error POM for this set of states is given by the square root measurement \cite{ban,eld}, and can be written $\hat{\Pi}_{i}^{ME} = \frac{2}{3} \ket{\phi_{i}^{ME}} \bra{\phi_{i}^{ME}}$ where
\begin{equation}
\begin{array}{ccl}
\ket{\phi_{0}^{ME}} &=& \frac{1}{\sqrt{2}} (\ket{0} + \ket{1}) \\
\ket{\phi_{1}^{ME}} &=& \frac{1}{\sqrt{2}} (\ket{0} + e^{2 \pi i /3} \ket{1}) \\
\ket{\phi_{2}^{ME}} &=& \frac{1}{\sqrt{2}} (\ket{0} + e^{-2 \pi i /3} \ket{1}).
\end{array}
\end{equation}
If this measurement is performed and outcome $\omega_{j}$ is obtained, then the probability that the state was indeed $\hat{\rho}_{j}$ is $P(\hat{\rho}_{j} | \omega_{j}) = \frac{1}{3}(1+ \sin 2 \theta)$.  This is plotted for comparison purposes alongside the optimal value of $\frac{2}{3}$ as a function of $\theta$ in Fig. \ref{Pc}.  It can be seen that for all values of $\theta$, except where the two strategies coincide at $\theta = \frac{\pi}{4}$, the new measurement strategy gives a greater confidence than that found for the minimum error strategy.

Analytic expressions were given above for the operators describing this new strategy for an arbitrary set of states (Eqs. \ref{MCPOM} and \ref{MCPOMpure}).  In deriving these the ansatz in Eq. \ref{ansatz} was used.  The significance of this can be explained by reference to some general transformation described by the invertible operator $\hat{A}$, which transforms $\hat{\rho} \rightarrow \hat{A} \hat{\rho} \hat{A}^{\dagger}$.  Under this transformation, Eq. \ref{confidence} becomes
\begin{equation}
P(\hat{\rho}_{j}|\omega_{j}) = \frac{p_{j} {\rm Tr}(\hat{A} \hat{\rho}_{j} \hat{A}^{\dagger} \hat{\Pi}_{j}^{\prime})}{{\rm Tr}(\hat{A} \hat{\rho} \hat{A}^{\dagger} \hat{\Pi}_{j}^{\prime})}
\end{equation}
for some positive operator $\hat{\Pi}_{j}^{\prime}$.  It is clear that if we define $\hat{\Pi}_{j}^{\prime} = (\hat{A}^{\dagger})^{-1} \hat{\Pi}_{j} \hat{A}^{-1}$, this conditional probability is identical for the original system $\hat{\rho}$ and the transformed system $\hat{A} \hat{\rho} \hat{A}^{\dagger}$, under the action of the positive operators $\hat{\Pi}_{j}$ and $\hat{\Pi}_{j}^{\prime}$ respectively.  Furthermore, as $\hat{A}$ is invertible, this transformation describes a one-to-one mapping between operators on the original and transformed systems.  Thus if the operator achieving maximum confidence is known for one system, it is easy to find that for the other system, simply by applying the appropriate transformation.  The advantage of the transformation used in Eq. \ref{ansatz} is that the operator $\hat{\Pi}_{j}^{\prime} \propto \hat{Q}_{j}$ which maximises this figure of merit for the transformed set $\{ \hat{\rho}_{j}^{\prime} \}$ is easily found.

The type of transformation discussed above can be realised as the result of a measurement associated with the POM $\{ \hat{A}^{\dagger} \hat{A}, \hat {\rm I} - \hat{A}^{\dagger} \hat{A} \}$.  Thus the measurement described by the probability operators in Eq. \ref{transform} can be viewed as a two step process.  In the first step, a measurement is performed with outcomes $\omega_{succ}$, $\omega_{fail}$, and associated POM elements
\begin{equation}
\hat{\Pi}_{succ} = \frac{p_{succ}}{D} \hat{\rho}^{-1}, \quad
\hat{\Pi}_{fail} = \hat {\rm I} - \hat{\Pi}_{succ},
\end{equation}
where $D$ is the dimension of the state space of the system, and $p_{succ}$ is the probability of occurrence of outcome $\omega_{succ}$.  To ensure positivity of both $\hat{\Pi}_{succ}$ and $\hat{\Pi}_{fail}$, $p_{succ}$ must satisfy the condition $0 \leq p_{succ} \leq \alpha_{i} D$ for all $i$, where $\alpha_{i}$ are the eigenvalues of $\hat{\rho}$.

When this step is performed, any given input state $\rho_{j}$ is transformed to $\hat{\rho}_{j}^{\prime}$, defined as above, with probability $p_{succ} {\rm Tr}(\hat{\rho}_{j} \hat{\rho}^{-1}) /D$ (corresponding to outcome $\omega_{succ}$).  This measurement strategy does not require that the operators in Eq. \ref{transform} form a complete measurement, and if outcome $\omega_{fail}$ is obtained, no further measurement is made, and the result may be interpreted as inconclusive.

The information provided by knowledge of the measurement outcome $\omega_{succ}$ also causes the associated probability distribution to be modified as follows
\begin{equation}
p_{j}^{\prime} = P(\hat{\rho}_{j} |\omega_{succ}) = \frac{P(\hat{\rho}_{j}) P(\omega_{succ}|\hat{\rho}_{j})}{P(\omega_{succ})} = \frac{p_{j}}{D} {\rm Tr}(\hat{\rho}_{j} \hat{\rho}^{-1}).
\end{equation}
It is easily verified that
\begin{equation}
\sum_{i} p_{i}^{\prime} = 1, \quad
\sum_{i} p_{i}^{\prime} \hat{\rho}_{i}^{\prime} = \frac{1}{D} \hat {\rm I} = \hat{\rho}^{\prime}.
\end{equation}

Note that the operators which give maximum confidence for this new set are immediately clear as the operator $\hat{\rho}_{j}^{\prime}$ describing any given state commutes with that describing the other states in the set $\hat{\rho}^{\prime} - p_{j} \hat{\rho}_{j}^{\prime}$.  Thus these two operators are simultaneously diagonalisable.  Also as $\hat{\rho}^{\prime} \propto \hat {\rm I}$, the same eigenvector corresponds to the largest eigenvalue of $\rho_{j}^{\prime}$ and the smallest eigenvalue of $\hat{\rho}^{\prime} - p_{j} \hat{\rho}_{j}^{\prime}$.  The optimum probability operator $\hat{\Pi}_{j}^{\prime}$ is a projection onto this eigenvector (Eq. \ref{Pj}).

The measurement described by operators $\{ \hat{\Pi}_{j}^{\prime} = \frac{c_{j} D}{p_{succ}} \hat{Q}_{j} \}$ is thus the second step in the process.  The probability of obtaining result $\omega_{j}$ when the system is in any given input state $\hat{\rho}_{i}$ can be written
\begin{equation}
\begin{array}{ccl}
P(\omega_{j} | \hat{\rho}_{i}) &=& P(\omega_{succ} | \hat{\rho}_{i}) P(\omega_{j} | \omega_{succ}, \hat{\rho}_{i}) \\
&=& \frac{p_{succ} {\rm Tr}(\hat{\rho}_{i} \hat{\rho}^{-1})}{D} {\rm Tr}(\hat{\rho}_{i}^{\prime} \frac{c_{j} D}{p_{succ}} \hat{Q}_{j}) \\
&=& {\rm Tr}(\hat{\rho}_{i} c_{j} \hat{\rho}^{-1/2} \hat{Q}_{j} \hat{\rho}^{-1/2}) = {\rm Tr}(\hat{\rho}_{i} \hat{\Pi}_{j}).
\end{array}
\end{equation}
Thus this two-step process is equivalent to the single step measurement described by probability operators $\{ \hat{\Pi}_{j}, \hat{\Pi}_{fail} \}$.  From the discussion above we know that $\hat{Q}_{j}$ is a projector onto the eigenstate of $\hat{\rho}_{j}^{\prime}$ with the largest eigenvalue.  For pure states, as $\hat{Q}_{j} = \hat{\rho}_{j}^{\prime}$, it is possible to choose the constants of proportionality such that $\hat{\Pi}_{j}^{\prime} = p_{j}^{\prime} D \hat{\rho}_{j}^{\prime}$ and these probability operators form a complete measurement.  For mixed states this is not possible, and the inconclusive outcome will have an additional component.  Thus for pure states the entire process may be interpreted as a projection of the initial states $\{ \hat{\rho}_{i} \}$ to the transformed set $\{ \hat{\rho}_{i}^{\prime} \}$, followed by a measurement along these states.

What are the properties of the transformed set?  As $\hat{\rho}^{\prime} = \frac{1}{D} \hat {\rm I}$, the states span the entire state space.  They are `maximally orthogonal' in the sense that the pure state with which any given state $\rho_{j}^{\prime}$ has largest overlap is also that with which the average of the remaining states $\hat{\rho}^{\prime} - p_{j}^{\prime} \hat{\rho}_{j}^{\prime}$ has smallest overlap.  In particular, for linearly independent sets, for which this strategy coincides with that of unambiguous discrimination, the initial states are projected onto mutually orthogonal states between which perfect discrimination is possible.  This is exactly the way in which unambiguous discrimination between two non-orthogonal states has been realised experimentally \cite{hut,cla}.  For linearly dependent sets, the above may be made clearer by reference to a qubit system.  The property $\hat{\rho}^{\prime} = \frac{1}{2} \hat {\rm I}$ means that the 3-D vector representing the state $\hat{\rho}_{j}^{\prime}$ on the Bloch sphere points in the opposite direction to that representing $\hat{\rho}^{\prime} - p_{j}^{\prime} \hat{\rho}_{j}^{\prime}$.

Thus we have constructed a measurement which allows us to be as confident as possible that when a measurement outcome leads us to identify a particular state, that state was indeed present.  As different outcomes are treated independently, an inconclusive outcome is sometimes necessary in order to form a physically realisable measurement.  We have given analytic expressions for the operators describing this optimal measurement for an arbitrary set of initial states, and have interpreted these expressions in terms of a two-step measurement process.  We have illustrated the new strategy by means of an example, and shown that for the set of states considered, when a state is identified, the probability that it was actually present is improved over the minimum error strategy.  This strategy is analogous to unambiguous discrimination, but is applicable to linearly dependent states.  We plan to demonstrate this strategy experimentally for the example considered here using optical polarisation.

\begin{acknowledgments}
This work was supported by the Universities of Glasgow and Strathclyde through a Synergy postgraduate studentship and by the Royal Society.
\end{acknowledgments}

\bibliography{MaxConf}

\end{document}